\documentclass[12pt]{article}
\usepackage{epsf,epsfig}

\newcommand{\autor}[1] {\begin{center}{\bf \lineskip .3cm #1} \end{center}}
\newcommand{\address}[1] {\begin{center}  {\normalsize \bf \it #1 }\end{center}}
\def\simgt{\rlap{\lower 3.5 pt \hbox{$\mathchar \sim$}} \raise 1pt \hbox {$>$}}
\def\simlt{\rlap{\lower 3.5 pt \hbox{$\mathchar \sim$}} \raise 1pt \hbox {$<$}}

\def\3half{\textstyle\frac32}

\newcommand{\gev}{\ {\rm GeV}}
\newcommand{\mev}{\ {\rm MeV}}
\newcommand{\lan}{{\langle}}
\newcommand{\ran}{{\rangle}}

\newcommand{\prd}{Phys. Rev. D }
\newcommand{\prl}{Phys. Rev. Lett. }
\newcommand{\plb}{Phys. Lett. B }
\newcommand{\jpg}{J. Phys. G }
\newcommand{\npb}{Nucl. Phys. B }
\newcommand{\zpc}{Z. Phys. C }
\newcommand{\epjc}{Eur. Phys. J. C }

\newcommand\be{\begin{equation}}
\newcommand\ee{\end{equation}}
\newcommand\bea{\begin{eqnarray}}
\newcommand\eea{\end{eqnarray}}

\def\be{\begin{eqnarray}}
\def\en{\end{eqnarray}}
\def\non{\nonumber}

\def\simgt{\rlap{\lower 3.5 pt \hbox{$\mathchar \sim$}} \raise 1pt \hbox {$>$}}
\def\simlt{\rlap{\lower 3.5 pt \hbox{$\mathchar \sim$}} \raise 1pt \hbox {$<$}}

\def\3half{\textstyle\frac32}

\def\simgt{\rlap{\lower 3.5 pt \hbox{$\mathchar \sim$}} \raise 1pt \hbox {$>$}}
\def\simlt{\rlap{\lower 3.5 pt \hbox{$\mathchar \sim$}} \raise 1pt \hbox {$<$}}

\def\3half{\textstyle\frac32}
\begin{document}
\begin{titlepage}
\vspace{1.0cm}
\begin{center}
\large\bf\boldmath Systematic Study on QCD Interactions of Heavy
Mesons with $\rho$ Meson \unboldmath
\end{center}
\vspace{0.8cm}\autor{Zuo-Hong Li,$^{1,~2,}~$\footnote{Corresponding
author.}~\footnote{E-mail: lizh@ytu.edu.cn} ~Wei Liu $^{1}$ and
Hai-Yan Liu $^{1}$}\vspace{0.7cm}
\address{$^1$ Department of Physics, Yantai University,
Yantai 264005, China\,\footnote{Mailing address}}
\address{$2$ CCAST (World Laboratory), P.O.Box 8730, Beijing 100080,
China} \vspace{1.0cm}
\begin{abstract}

The strong interactions of the negative-parity heavy mesons with
$\rho $ meson may be described consistently in the context of an
effective lagrangian, which is invariant under isospin $SU(2)$
transformation. Four coupling constants $g_{HH\rho}$,
$f_{H^*H\rho}$, $g_{H^*H^*\rho}$ and $f_{H^*H^*\rho}$ enter the
effective lagrangian, where $H$ $(H^*)$ denotes a pseudoscalar
bottom or charm meson (the corresponding vector meson). Using QCD
light cone sum rule (LCSR) method and, as inputs, the hadronic
parameters updated recently, we give an estimate of $g_{H^*H^*\rho}$
and $f_{H^*H^*\rho}$, about which little was known before, and
present an improved result for $g_{HH\rho}$ and $f_{H^*H\rho}$.
Also, we examine the heavy quark asymptotic behavior of these
nonperturbative quantities and assess the two low energy parameters
$\beta$ and $\lambda$ of the corresponding effective chiral
lagrangian.

\end{abstract}
~~~~~~~~{\it PACS numbers:} 13.75.Lb, 12.38.Lg, 14.40.Nd, 14.40 Cs

~~~{\it Keywords:~~Strong Couplings, Light Cone Sum Rules, Heavy
Mesons, $\rho$ Meson}

\end{titlepage}

\section{Introduction}
~~~~At present, we have the two well-established theoretical
frameworks for describing a large class of two body hadronic decays
of B mesons, that is, QCD factorization (QCDF) \cite{BBNS} approach
and soft collinear effective theory (SCET) \cite{SCET}.
Long-distance parameters enter inevitably, however, as important
inputs in their phenomenological applications. One is yet confronted
with the difficult task to cope with the nonperturbative problems.
Numerous theoretical works are devoted to this subject. Among all
the existing nonperturbative approaches, QCD light cone sum rules
(LCSR's) \cite{LCSR, LCSRYK} have proved to be particularly
powerful. This is due to the facts: (1) Contrary to conventional sum
rule calculations \cite{SVZ} on the form factors for heavy to light
decays, LCSR results turn out to be consistent in the heavy quark
limit $m_Q\rightarrow\infty$ \cite{Ali}; (2) LCSR approach allows us
to consistently explore the dependence of the form factors on the
momentum transfer $q^2$ in the whole kinematically accessible range
[7--10], by combing its results, which are valid for small and
intermediate $q^2$, with the pole model description for the form
factors at large $q^2$. The LCSR method has extensively been applied
to study the semileptonic [4, 6--11, 12] and hadronic \cite{LCSR5}
decays of heavy mesons. A recent LCSR reanalysis of heavy-to-light
transitions can be found in \cite{LCSR4}.

Along with the great progresses in the experiment on $B$ physics, we
are confronting a more formidable challenge to deal with the
nonperturbative dynamics. The data accumulated at B factories and
CLEO give a hint that there may be large contributions from the
final state interactions (FSI's), which are typically
nonperturbative, in some of two body hadronic B decays. In the
absence of a rigorous approach to FSI's, one may either resort to
Regge theory \cite{Regge1} to estimate their effects \cite{Regge2},
or mimic them by the soft rescattering of two intermediate particles
so that they could be viewed as a one particle exchange process
calculable at the hadron level. In comparison, the latter is of more
intuitive physical picture and thus is accepted more readily.
Employing the one particle exchange model, calculations of the FSI's
in both B and D decays have been undertaken many a time in the
literature [16--18]. For a recent application of this approach, ones
are referred to \cite{FSI2, FSI3}. A precondition of doing such a
calculation, however, is that the related couplings are supposed to
be known, which parameterize the strong interactions among the
underlying meson fields. The most interesting is the situation that
heavy mesons interact with a light meson, where the corresponding
couplings are also crucial to determine the normalization of heavy
to light form factors at large momentum transfer in the pole
dominance models [7--10]. These couplings have to be assessed
adopting a certain phenomenological method, except that few of them
can be extracted directly from the experimental data. In the case
where the light meson concerned is a pseudoscalar meson, the related
couplings have undergone a systematic investigation, in the
frameworks of both LCSR's \cite{LCSR1, LCSRLI1, PILCSR} and
conventional sum rules \cite{PISR}. In contrast, the existing
discussion is incomplete about the interactions of heavy mesons with
a light vector meson, despite some efforts being made [17, 21--23].

The strong interactions can be described between the negative-parity
heavy mesons and $\rho$ meson by constructing an effective
lagrangian which respects the SU(2)symmetry in the isospin space.
Letting us define an isospin doublet $B$ composed of the
pseudoscalar bottom meson fields $B^+$ and $B^0$ and the
corresponding vector doublet $B^*_{\mu}$:
\begin{displaymath}
B={B^+\choose B^0},~~~~~ B^*_\mu={B^{*+}_{\mu}\choose B^{*0}_{\mu}},
\end{displaymath}
with the hermitian conjugate forms
\begin{displaymath}
B^{\dagger}=(B^{-}\quad
\bar{B}^{0}),~~~~~B^{*\dagger}_\mu=(B^{*-}_{\mu}\quad
\bar{B}^{*0}_{\mu}),
\end{displaymath}
and representing the isospin triplet of the $\rho$ meson field by
\begin{displaymath}
{P}_\mu=\left(\begin{array}{ccc}
\rho_\mu^0&~~\sqrt{2}\rho^+_\mu\\
\sqrt{2}\rho^-_\mu&~~-\rho_\mu^0\end{array}\right),
\end{displaymath}
we can build the effective lagrangian of the required symmetry as
\begin{eqnarray}
{\cal L}&=& ig_{BB\rho}Tr\left [\left(
 B^\dagger\buildrel\leftrightarrow\over\partial_\mu
 B\right)P^\mu\right]-2f_{B^*B\rho}\varepsilon^{\mu\nu\alpha\beta}Tr\left[\left(B^\dag\buildrel\leftrightarrow\over{\partial_\mu}B^{*}_\nu-
 B^{*\dag}_\nu\buildrel\leftrightarrow\over{\partial_\mu}
 B\right)\partial_\alpha
 P_\beta\right]\nonumber \\
 &+&ig_{B^*B^*\rho}Tr\left [\left(\bar
 B^{*\dagger}_\mu\buildrel\leftrightarrow\over\partial_\nu
 B^{*\mu}\right)P^\nu\right]+4if_{B^*B^*\rho}m_{B^*}Tr\left [\left(
 B^{*\dagger}_{\mu}B^*_{\nu}\right)\left(\partial^{\mu}P^{\nu}-\partial^{\nu}P^{\mu}\right)\right]\label{1}
 \end{eqnarray}
and analogous one for the charm mesons. In the established effective
lagrangian, four coupling constants $g_{BB\rho}$, $f_{B^*B\rho}$,
$g_{B^*B^*\rho}$ and $f_{B^*B^*\rho}$ are introduced, as a result of
$SU(2)$ isospin asymmetry, to describe the strength of the strong
interactions among the related meson fields. Whereas $g_{BB\rho}$
and $f_{B^*B\rho}$ serve as describing the $B$-$B$-$\rho$ and
$B^*$-$B$-$\rho$ interactions respectively, the other two as
characterizing the $B^*$-$B^*$-$\rho$ interactions. As explained
clearly later, these couplings are of definite physical meaning and
in the limit $m_Q\rightarrow\infty$, they coincide, up to a
prefactor, with one of the two low energy parameters $\beta$ and
$\lambda$, which parameterize the effective chiral lagrangian for
the heavy mesons and light vector resonances \cite{CHIRAL}. Such
that the effective description formulated in (1) is consistent with
the effective chiral lagrangian approach. In term of these couplings
the relevant hadronic matrix elements are parameterized as:
 \be \langle \bar
B^{0}(p)\rho^-(q,\epsilon)|i{\cal
L}|B^{-}(p+q)\rangle=2\sqrt{2}ig_{BB\rho}p\cdot\epsilon^*,
\label{2}\en \be \langle \bar
B^{*0}(p,\eta)\rho^-(q,\epsilon)|i{\cal
L}|B^{-}(p+q)\rangle=-4\sqrt{2}if_{B^*B\rho}\epsilon_{\mu\alpha\beta\gamma}\eta^{\ast\mu}q^{\alpha}
\epsilon^{\ast \beta}p^{\gamma}, \label{3}\en
 \be
 \langle \bar B^{*0}(p,\eta)\rho^-(q,\epsilon)|i{\cal L}&&\!\!\!|B^{*-}(p+q,\xi)\rangle = -2\sqrt{2}ig_{B^*B^*\rho}\left(\eta^*\cdot\xi\right)
\left(p\cdot\epsilon^*\right)\nonumber\\&&-4\sqrt{2}if_{B^*B^*\rho}m_{B^*}\left[\left(\eta^*\cdot\epsilon^*\right)\left(\xi\cdot
q\right)-\left(\xi\cdot\epsilon^*\right)\left(\eta^*\cdot
q\right)\right],\label{4}
  \en
where the momentum and polarization vector assignment is specified
in brackets. These hadronic matrix elements, as those parameterizing
the strong interaction processes of heavy mesons and a pionic meson,
would play a prominent role in the phenomenological study of heavy
flavor physics. As aforementioned, however, for the time being we
are devoid of an all-around knowledge of them. The previous LCSR
calculation is just confined to the case of $g_{BB\rho}$ and
$f_{B^*B\rho}$ \cite{LCSRLI2}, and the effective parameters $\beta$
and $\lambda$ are merely investigated on the basis of the vector
dominance assumption \cite{FSI2, CHIRAL}. On the other hand, the
existing LCSR results call for a recalculation with an updated
hadronic parameter.

In this letter, we intend to give a LCSR estimate of
$g_{B^*B^*\rho}$ $(g_{D^*D^*\rho})$ and $f_{B^*B^*\rho}$
$(f_{D^*D^*\rho})$, along with an improved numerical prediction of
$g_{BB\rho}$ $(g_{DD\rho})$ and $f_{B^*B\rho}$ $(f_{D^*D\rho})$, and
then make an investigation into $m_Q$ scaling behavior of the
resultant sum rules, present our LCSR results for the effective
parameters $\beta$ and $\lambda$.

\section{LCSR Calculation on Strong Couplings}
~~~~~~We focus on the bottom case and begin with a discussion of the
$B^*B^*\rho$ coupling. For implementing a QCD LCSR calculation on
$g_{B^*B^*\rho}$ and $f_{B^*B^*\rho}$, it is advisable to make use
of the following correlation function:
\begin{eqnarray}
H_{\mu\nu}\left(p,~q,~e\right)&=&i\int d^4xe^{ipx}\langle\rho^-(q,\epsilon)|T\bar{d}(x)\gamma_\mu b(x),\bar{b}(0)\gamma_\nu u(0)|0\rangle\non\\
&=&H\left( p^2,\left(
p+q\right)^2\right)g_{\mu\nu}p\cdot\epsilon^*+\widetilde{H}\left(
p^2,\left(
p+q\right)^2\right)\left(q_\mu\epsilon_\nu^*-q_\nu\epsilon_\mu^*\right)
+\cdots,\label{5}
\end{eqnarray}
where ellipses indicate the remaining Lorentz structures. The
hadronic form of the correlation function (5) is easily obtained by
saturating it with a complete set of intermediate states with the
same quantum numbers as the interpolating current operators.
However, we need to do it with care, for the vector current
operators can also couple with the set of scalar bottom meson with
positive-parity, besides that of vector bottom meson. On taking into
account all the possible hadronic contributions to
$H_{\mu\nu}\left(p,~q,~e\right)$, we find that the invariant
functions $H$ and $\widetilde{H}$ receive only the contributions
from the set of vector bottom meson. Isolating the pole contribution
of the lowest $B^*$ meson and parameterizing these from the higher
states in a form of double dispersion integral
starting with the threshold $s_{0}$, we have the desired hadronic forms for $H(p^2,(p+q)^2)$ and $\widetilde{H}(p^2,(p+q)^2)$:
\begin{eqnarray}
H^h\left(p^2,\left(p+q\right)^2\right)&=&
\frac{-2\sqrt{2}m^2_{B^*}f^2_{B^*}g_{B^*B^*\rho}}{(p^2-m^2_{B^*})[(p+q)^2-m^2_{B^*}]}+\int\int\frac{\rho^h\left(s_1,s_2\right)ds_1ds_2}{\left(s_1-p^2\right)\left[s_2-(p+q)^2\right]},
\label{10}\end{eqnarray}

\begin{eqnarray}
\widetilde{H}^h\left(p^2,\left(p+q\right)^2\right)&=&
\frac{4\sqrt{2}m^3_{B^*}f^2_{B^*}f_{B^*B^*\rho}}{(p^2-m^2_{B^*})[(p+q)^2-m^2_{B^*}]}+\int\int\frac{\tilde{\rho}^h\left(s_1,s_2\right)ds_1ds_2}{\left(s_1-p^2\right)\left[s_2-(p+q)^2\right]},
\label{11}\end{eqnarray}
with $f_{B^*}$, as defined usually, being
the decay constant of $B^*$ meson and $\rho^h$ $(\tilde{\rho}^h)$
the hadron spectral function.

QCD calculation of the correlator (5) can be carried out for the
negative and large values of $p^{2}-m^2_Q$ and
$\left(p+q\right)^{2}-m^2_Q$, which render the operator product
expansion (OPE) valid near the light-cone $x^{2}=0$. Since the
underlying heavy quark is sufficiently far off shell in the
kinematical regions, in terms of the light-cone expansion the soft
gluon emissions from the heavy quark contribute just a higher twist
effect, which is concerned with the quark-antiquark-gluon $(q\bar q
g)$ components of the $\rho$ meson distribution amplitudes. As
verified by the numerous LCSR calculations, omitting the gluon
emission contributions may be considered a better approximation. For
the present calculation, we will use the free $b$ quark propagator:
\begin{equation}
\lan 0|Tb\left( x\right) \overline{b}\left( 0\right)|0\ran
=\frac{1}{\left( 2\pi \right) ^{4}i}\int d^{4}ke^{-ik\cdot
x}\frac{k\!\!\!/+m_{b}}{m_{b}^{2}-k^{2}}. \label{7} \label{12}
\end{equation}
Substituting (8) in (5) and using the $\gamma$ algebraic relations
\begin{equation}
\sigma_{\mu\nu}=i\left(\gamma_\mu\gamma_\nu-g_{\mu\nu}\right)
\end{equation}
and
\begin{equation}
\gamma_\mu\sigma_{\rho\lambda}=i(g_{\mu\rho}\gamma_\lambda-g_{\mu\lambda}\gamma_\rho)
+\varepsilon_{\mu\rho\lambda\theta}\gamma^\theta\gamma_5,
\end{equation}
we are led to the light cone wavefunctions of the $\rho$ meson
defined by \cite{WF3, WF4, WF5}
\begin{eqnarray}
\left\langle \rho \left(q,e\right) \left|\bar{d}\left( x\right)
\gamma _{\mu }u\left( 0\right)\right|0\right\rangle &=&f_{\rho
}m_{\rho }\left\{ \frac{e^{\left( \lambda \right) \ast }\cdot
x}{q\cdot x}q_{\mu }\int_{0}^{1}due^{iuq\cdot x}\left[ \varphi
_{\parallel }\left( u,\mu \right) +\frac{m_{\rho
}^{2}x^{2}}{16}A\left( u,\mu \right) \right] \right.
\nonumber \\
&+&\left( e_{\mu }^{\left( \lambda \right)\ast }-q_{\mu
}\frac{e^{\left( \lambda \right) \ast }\cdot x}{q\cdot x}\right)
\int_{0}^{1}due^{iuq\cdot x}g_{\perp
}^{\left( v\right) }\left( u,\mu \right) \nonumber \\
&-&\left.\frac{1}{2}x_{\mu }\frac{e^{\left( \lambda \right)\ast }\cdot x}{%
\left( q\cdot x\right) ^{2}}m_{\rho }^{2}\int_{0}^{1}due^{iuq\cdot
x}C\left( u,\mu \right) \right\},  \label{16}
\end{eqnarray}
\begin{eqnarray}
\left\langle \rho(q,e)\left|\bar{d}\left( x\right) u\left( 0\right)
\right|0\right\rangle=-i/2f^T_{\rho}m^2_{\rho}(e^{\lambda}\cdot
x)\int^1_0 du e^{iuq\cdot x}h^s_{\parallel}(u,\mu_b), \label{17}
\end{eqnarray}
\begin{eqnarray}
\left\langle \rho(q,e) \left|\bar{d}\left( x\right) \sigma ^{\alpha
\beta}u\left(0\right) \right|0\right\rangle &=&-if_{\rho
}^{T}\left\{ \left( e_{\left( \lambda \right) }^{\ast \alpha
}q^{\beta }-e_{\left( \lambda \right) }^{\ast \beta }q^{\alpha
}\right) \int_{0}^{1}due^{iuq\cdot x}\left[ \varphi _{\perp}\left(
u\right)
+\frac{1}{16}m_{\rho }^{2}x^{2}A_{T}\left(u\right) \right] \right.\nonumber \\
&+&\left( q^{\alpha }x^{\beta }-q^{\beta }x^{\alpha }\right)
\frac{e_{\left( \lambda \right) }^{\ast }\cdot x}{\left( q\cdot
x\right) ^{2}}m_{\rho
}^{2}\int_{0}^{1}due^{iuq\cdot x}B_{T}\left( u\right)  \nonumber \\
&+&\left.\frac{1}{2}\left( e_{\left( \lambda \right) }^{\ast \alpha
}x^{\beta }-e_{\left( \lambda \right) }^{\ast \beta }x^{\alpha
}\right) \frac{m_{\rho }^{2}}{q\cdot x}\int_{0}^{1}due^{iuq\cdot
x}C_{T}\left( u\right)\right\},\label{18}
\end{eqnarray}
\begin{eqnarray}
\left\langle \rho(q,e)\left|\bar{d}\left( x\right) \gamma _{\mu
}\gamma _{5}u\left( 0\right) \right|
0\right\rangle=\frac{1}{4}f_{\rho}m_{\rho}\varepsilon_{ \mu\alpha
\beta\gamma }q^{\alpha}e^{\beta}x^{\gamma}\int^1_0due^{iuq\cdot
x}g^{(a)}_{\perp}(u,\mu_{b}),\label{19}
\end{eqnarray}
where $f_{\rho}$ stands for the usual decay constant of the $\rho$
meson, and $f_{\rho}^T$ is defined as
$\left\langle0|\bar{u}\sigma_{\mu\nu}d|
\rho\right\rangle=if_{\rho}^T\left(e_{\mu}^{
(\lambda)}q_{\nu}-e_{\nu}^{(\lambda)} q_{\mu}\right)$; both $\varphi
_{\parallel }\left( u,\mu \right)$ and $\varphi _{\perp}\left( u,\mu
\right)$ denote the leading twist-2 distribution amplitudes,
$g_{\perp }^{\left( v\right) }\left( u,\mu \right) $, $g_{\perp
}^{\left( a\right) }\left( u,\mu \right)$ and
$h^s_{\parallel}(u,\mu_b)$ refer to the twist-3 ones, and the others
are all of twist-4. With all these expressions, a straightforward
calculation yields the following QCD forms for $H(p^2,(p+q)^2)$ and
$\widetilde{H}(p^2,(p+q)^2)$:
\begin{eqnarray}
H^{QCD}\left(p^2,\left(p+q\right)^2 \right)&=&-\left\{f_{\rho
}m_{\rho }\int_{0}^{1}du \frac{\varphi _{\parallel }\left( u\right)
}{m_{b}^{2}-\left( p+uq\right)^{2}}
+f_{\rho}^Tm_{\rho}^2m_b\int_0^1du\frac{h^{(s)}_{\parallel}(u,\mu_b)}{\left[
m_{b}^{2}
-\left(p+uq\right)^{2}\right] ^{2}}\right.\non\\
&&\left.-\frac{1}{2}f_{\rho }m_{\rho
}^{3}\int_{0}^{1}du\left[\frac{A\left( u\right)}{\left[
m_{b}^{2}-\left( p+uq\right) ^{2}\right]
^{2}}+\frac{m_b^2\left(A\left( u\right)+8\widetilde{C}\left(
u\right)\right)}{\left[ m_{b}^{2}-\left( p+uq\right) ^{2}\right]
^{3}}\right] \right\},\label{22}
\end{eqnarray}
\begin{eqnarray}
\widetilde{H}^{QCD}\left(p^2,\left(p+q\right)^2 \right)&=&m_bf_{\rho
}^{T} \int_{0}^{1}du \frac{\varphi _{\perp}\left(
u\right)}{m_{b}^{2}-\left(p+uq\right) ^{2}}\non\\
&&+\frac{1}{4}f_\rho
m_\rho\int_{0}^{1}du\left[\frac{g_{\perp}^{(a)}(u)+4\left(u\frac{dg_\perp^{(a)}}{\!\!\!\!\!du}(u)+g_{\perp}^{(v)}(u)\right)}{m_{b}^{2}-\left(
p+uq\right) ^{2}}+\frac{2m^2_bg_\perp^{(a)}(u)}{\left[
m_{b}^{2}-\left( p+uq\right) ^{2}\right]
^{2}}\right]\non\\
&&-f_\rho^T
m_bm_\rho^2\int_{0}^{1}du\left[\frac{m_b^2A_T(u)}{2\left[
m_{b}^{2}-\left(
p+uq\right)^{2}\right]^{3}}+\frac{2\widetilde{B}_T(u)+u\bar{C}_T(u)}{\left[m_{b}^{2}-\left(
p+uq\right) ^{2}\right]^2}\right]\non\\
&&+\frac{1}{4}f_{\rho}
m_\rho^3\int_{0}^{1}du\left[\frac{\bar{A}(u)-8u\widetilde{C}(u)}{\left[m_{b}^{2}-\left(
p+uq\right)
^{2}\right]^2}+\frac{2m_b^2\bar{A}(u)}{\left[m_{b}^{2}-\left(
p+uq\right) ^{2}\right]^3}\right].\label{23}
\end{eqnarray}
In the derivations of (15) and (16) we have introduced two auxiliary
functions $\bar{f}(u)=\int _0^uf(v)dv$ and
$\widetilde{f}\left(u\right) =\int_{0}^{u} \bar{f} \left(
v\right)dv$. It is needed to convert both the QCD expressions into a
form of double dispersion integral, for matching them onto the
individual hadronic forms. However, we note that it is sufficient to
do it only for the twist-2 and-3 parts. The relevant QCD spectral
densities are easily obtained with the standard method \cite{SR1}.
To this end, the following formula is useful:
\begin{eqnarray}
\widehat{B}_{M_{1}^{2}}
\widehat{B}_{M_{2}^2}\frac{(l-1)!}{[m_{b}^{2}-\left(
p+uq\right)^{2}]^l}=\frac{(M^{2})^{2-l}}{M_{1}^{2}M_{2}^{2}}e^{-1/M^2\left[
m_{b}^{2}+m_{\rho }^{2}u_{0}(1-u_{0})\right]}\delta(u-u_0),
\label{24}
\end{eqnarray}
where $\widehat{B}_{M_{1}^{2}}$ and $\widehat{B}_{M_{2}^2}$ are the
Borel operators, the Borel parameters $M_1^2$ and $M^2_2$ are
associated with $p^2$ and $(p+q)^2$ respectively, $M^{2}=
M_{1}^{2}M_{2}^{2}/(M_{1}^{2}+M_{2}^{2})$ and
$u_{0}=M_{1}^{2}/\left( M_{1}^{2}+M_{2}^{2}\right) $.
Simultaneously, we can set $M_{1}^{2}=M_{2}^{2}$, because of the
symmetry of the correlator, so that the distribution amplitudes
entering the QCD spectral densities take only their values at the
symmetry point $u_{0}=1/2$. Here we omit the final expressions for
the QCD spectral functions to save some spaces.

To proceed, we perform the double Borel transformation
$-p^2\rightarrow M_1^2$, $-(p+q)^2\rightarrow M^2_2$ for both the
hadronic and QCD representations. The use of the quark-hadron
duality results in the final sum rules for the products
$f_{B^*}^{2}g_{B^*B^*\rho}$ and $f_{B^*}^{2}f_{B^{\ast }B^*\rho}$:
\begin{eqnarray}
f_{B^*}^{2}g_{B^*B^*\rho
}&=&\frac{\sqrt{2}}{4m_{B^*}^{2}}e^{\frac{m_{B^*}^{2}}{M^{2}}}
\left\{f_{\rho}m_{\rho}M^2\left[e^{-\frac{1}{M^2}
\left(m_b^2+\frac{1}{4}m_{\rho}^2\right)}-e^{-\frac{s_0}{M^2}}\right]
\varphi_{\parallel}\left(1/2\right)\right.\nonumber \\
&&\left.+~\frac{1}{4}m_{\rho }^{2}e^{-\frac{1}{M^{2}}\left(
m_{b}^{2}+\frac{1}{4}m_{\rho }^{2}\right)}\left[4m_{b}f_{\rho
}^{T}h_{\parallel }^{\left( s\right)}\left(1/2\right)-f_{\rho
}m_{\rho } \left(
1+\frac{m_{b}^2}{M^{2}}\right)A\left(1/2\right)\right.\right.\nonumber \\
&&\left.\left.-~\frac{8f_\rho m_\rho
m_{b}^2}{M^{2}}\widetilde{C}\left(1/2\right)\right] \right\},
\label{26}
\end{eqnarray}
\begin{eqnarray}
f_{B^{\ast }}^2f_{B^{\ast }B^*\rho }&=&\frac{\sqrt{2}}{8
m^3_{B^{\ast }}}e^{\frac{m_{B^{\ast }}^2}{{M}^{2}} }\left\{M^2\left(
e^{-\frac{1}{{M}^{2}}\left( m_{b}^{2}+\frac{1}{4} m_{\rho
}^{2}\right) }-e^{-\frac{s_0}{{M}^{2}}}\right)\right.\non\\&&
\times\left[f_\rho^Tm_b\varphi_{\perp}\left(1/2\right)+\frac{1}{8}f_\rho
m_\rho\left(2g_{\perp}^{(a)}\left(1/2\right)+\frac{dg_\perp^{(a)}}{\!\!\!du}\left(1/2\right)+4g_{\perp}^{(v)}\left(1/2\right)\right)\right]
\nonumber \\
&&+~\frac{1}{2}m_{\rho}e^{- \frac{1}{{M}^{2}}\left(
m_{b}^{2}+\frac{1}{4}m_{\rho }^{2}\right)}\left[f_\rho
m_b^2g^{(a)}_{\perp}\left(1/2\right)\right.\non\\&&-2f_\rho^Tm_bm_\rho\left(\frac{m_b^2}{4M^2}A_{T}
\left(1/2\right)+2\widetilde{B}_{T} \left(1/2\right)+\frac{1}{2}\bar
C_T\left(1/2\right)\right)\non\\&&\left.\left.+~f_\rho
m_\rho^2\left(\left(1+\frac{m_b^2}{2M^2}\right)\bar
A\left(1/2\right)-\frac{1}{2}\widetilde{C}\left(1/2\right)\right)\right]\right\}.
\label{27}
\end{eqnarray}

We proceed to the numerical computation of the sum rules. To
consistently specify the input, we take \cite{LCSR1}
$m_b=4.7\pm0.1\gev$, $m_{B^*}=5.325\gev$, $f_{B^{\ast}}=160\mev$ and
$s_0=35\pm 1\gev$ for the bottom channels. Some of the parameters
related to the $\rho $ meson are chosen as: $m_{\rho }=770\mev$,
$f_{\rho }=216\pm 3\mev$ and $f_{\rho }^{T}(\mu=1\gev)=165\pm 9\mev$
\cite{WF1}. The most important sources of uncertainty are the light
cone wavefunctions of $\rho$ meson. It is demonstrated that the
wavefunctions can be expanded in terms of matrix elements of
conformal operators. Based on this expansion, the first attempt was
made in \cite{WF2} to understand the twist-2 distribution amplitudes
of light vector mesons. Since a modified result was put forward
\cite{WF3} the model wavefunctions of light vector mesons, up to
twist-4, have undergone a successive examination and improvement
\cite{WF4, WF5, WF1}. Very recently, a more systematic inspection
was made of the existing model parameters and the updated results
were reported in \cite{WF1}. Here, we will make use of the findings
of \cite{WF1}, for the related distribution amplitudes of $\rho$
meson. Certainly, in the present applications the appropriate
normalization scale should be set at the typical virtuality of the
$b$ quark: $u_b=\sqrt{m_{B}^{2}-m_{b}^{2}}$. At this scale, the
numerical values of the nonperturbative quantities involved, which
contain the model parameters and $f_{\rho }^{T}$, can be reached by
use of the renormalization group equations. Using the inputs fixed,
the range of the Borel variable $M^2$ can be determined by demanding
that the 4-twist parts contribute less than $10\%$, while the higher
resonance and continuum contributions don't excess $30\%$. In both
the cases, the Borel interval to satisfy the above criteria is
$6\leq M^{2}\leq 12\gev^{2}$. From the sum rule "windows", it
follows that $f_{B^*}^{2}g_{B^*B^*\rho}=0.048\pm0.013\gev^2$ and
$f_{B^{\ast }}^2f_{B^{\ast }B^*\rho } =0.021\pm0.007\gev$. The
uncertainties quoted are in view of the variations of the $b$ quark
mass $m_b$, threshold $s_0$ and Borel parameter $M^2$. Dividing
these two sum rules by $f_{B^*}^2$ yields $g_{B^*B^*\rho}=1.88$,
$f_{B^{\ast }B^*\rho }=0.82\gev^{-1}$, where we give only the
central values of the numerical results.

With a definition different from the present ones by a constant
factor, the remaining two couplings $g_{BB\rho}$ and
$f_{B^{\ast}B\rho}$ have been computed in the same approach
\cite{LCSRLI2}. However, the numerical results are not
straightforwardly available for a consistent discussion, because
they are derived with the inputs, most of which, including the model
wavefunctions, are other than those used here and improved to a
certain extent. An updated estimate is obligatory. In passing, it is
deserving of mention that there is an unfortunate error checked out
by us in the previous LCSR calculation on the $B^{\ast}$-$B$-$\rho$
coupling (where the factor of $3/4$ in the term proportional
$A_T(1/2)$ should be modified as $-1/4$), but with a small numerical
impact. With this corresponding change, the LCSR expressions of the
present concern can be achieved trivially from (17) and (23) of
\cite{LCSRLI2}, for the products $f_B^2g_{BB\rho}$ and
$f_{B^{\ast}}f_B f_{B^{\ast}B\rho}$:
\begin{eqnarray}
f_{B}^{2}g_{BB\rho}
&=&\frac{\sqrt{2}m_{b}^{2}}{4m_{B}^{4}}e^{\frac{m_B^2}{M^2}}
\left\{f_{\rho}m_{\rho}M^2\left[e^{-\frac{1}{M^2}
\left(m_b^2+\frac{1}{4}m_{\rho}^2\right)}-e^{-\frac{s_0}{M^2}}\right]
\varphi_{\parallel}\left(1/2\right)\right.\nonumber\\
&&+\left.\frac{1}{4}m_{\rho }^{2}e^{-\frac{1}{M^{2}}\left(
m_{b}^{2}+\frac{1}{4}m_{\rho }^{2}\right)}\left[4m_{b}f_{\rho
}^{T}h_{\parallel }^{\left(
s\right)}\left(1/2\right)\right.\right.\nonumber\\
&&-\left.\left.f_{\rho}m_{\rho}\left(1+\frac{m_b
^2}{M^2}\right)\left(A\left(1/2\right)+8
\widetilde{C}\left(1/2\right)\right)\right]\right\},\label{24}
\end{eqnarray}
\begin{eqnarray}
f_{B^{\ast }}f_{B}f_{B^{\ast }B\rho }&=&\frac{\sqrt{2}m_{b}}{8
m_{B^{\ast }}m_{B}^{2}}e^{\frac{m_{B^{\ast }}^2+m_{B}^{2}}{2{M}^{2}}
}\left\{ f_{\rho }^{T}M^{2}\left( e^{-\frac{1}{{M}^{2}}\left(
m_{b}^{2}+\frac{1}{4} m_{\rho }^{2}\right)
}-e^{-\frac{s_0}{{M}^{2}}}\right)\varphi_{\perp}\left(1/2\right)\right.
\nonumber \\
&&+\left.\frac{1}{4}m_{\rho}e^{- \frac{1}{{M}^{2}}\left(
m_{b}^{2}+\frac{1}{4}m_{\rho
}^{2}\right)}\left[2m_bf_{\rho}g^{(a)}_{\perp}\left(1/2\right)\right.\right.\nonumber
\\&&-\left.\left. m_{\rho}f_{\rho }^{T}\left( 1+\frac{m_{b}^{2}}{{M}^{2}}\right)
A_{T}\left(1/2\right)\right]\right\}, \label{25}
\end{eqnarray}
where the two additional parameters, $m_B$ and $f_B$, for the bottom
meson channels are taken as $m_B=5.279\gev$ and $f_{B}=140\mev$. Our
observation is that these two sum rules can share a Borel range,
which is the about same as that for the $B^*B^*\rho$ case, and
provide the numerical predictions:
$f_{B}^{2}g_{BB\rho}=0.037\pm0.008\gev^2$ and $f_{B^{\ast
}}f_{B}f_{B^{\ast }B\rho }=0.019\pm0.005\gev$, from whose central
values we have $g_{BB\rho}=1.89$, $f_{B^{\ast
}B\rho}=0.85\gev^{-1}$.

A physical interpretation is in order on the LCSR predictions
presented above. As shown explicitly, there exist the approximate
sum rule relations $g_{BB\rho}\approx g_{B^*B^*\rho}$ and
$f_{B^{\ast }B\rho}\approx f_{B^{\ast }B^*\rho}$, for the coupling
constants appearing in the effective lagrangian (1). This may be
accounted for intuitively by observing the construction of the
effective lagrangian: The terms proportional to $g_{BB\rho}$ and
$g_{B^*B^*\rho}$ can be identified as describing the charge
interactions between the $B(B^*)$ and $\rho$ meson fields, while the
other two parts may be interpreted as indicating the magnetic
interactions of the underlying bottom mesons with $\rho$ meson
fields. It is not surprising, therefore, that the relations
$g_{BB\rho}= g_{B^*B^*\rho}$ and $f_{B^{\ast }B\rho}=f_{B^{\ast
}B^*\rho}$ should hold exactly in the limit
$m_Q\longrightarrow\infty$, because of the heavy quark spin
symmetry. In the following section, we are going to return to this
problem, putting it the test whether the LCSR calculations could
precisely give the asymptotic relations deduced from the heavy quark
spin symmetry.

Situations of the charm mesons can in parallel be discussed, using
the LCSR formulae (18)-(21) with a replacement of the corresponding
inputs. Of course, it is generally believed that in this case the
gluon emission corrections from the charm quarks may be relatively
important, due to the smaller heavy quark mass. Still, we omit them
for a consistent purpose. The parameters for the charm channels are
set as \cite{LCSR1}: $m_c=1.3\gev$, $m_D=1.87\gev$,
$m_{D^{\ast}}=2.01\gev$, $f_D=170\mev$, $f_{D^{\ast}}=240\mev$ and
$s_0=6\gev^2$. In addition, we need to set the proper scale at
$\mu_c=\surd\overline{m_D^2-m_c^2}$. Along the same line as in the
bottom case, the numerical analysis can be performed. Subject to an
evaluation of uncertainty, the yielded sum rule results for the
couplings are summarized as: $g_{DD\rho}=1.63$,
$g_{D^*D^*\rho}=1.68$, $f_{D^{\ast }D\rho }=0.81\gev^{-1}$, and
$f_{D^{\ast }D^*\rho }=0.78\gev^{-1}$.

We would like to compare the present LCSR predictions with the ones,
which are obtained with the inputs proposed earlier in \cite{WF3,
WF4} for the $\rho$ meson parameters, to see that to what extent
LCSR calculations have been improved with the updated parameters. It
is demonstrated that in the bottom case, using the updated
parameters can increase the LCSR evaluations by about $20\%$ and a
few percent, respectively, for the charge and magnetic interactions.
The corresponding changes in the charm case amount to an order of
$10\%$ and of few percent, respectively.

A further improvement on the LCSR results proposed here is expected,
since in the present case the $q\bar qg$ components of $\rho$ meson
don't enter in consideration, in addition to the QCD radiative
corrections, and a further update is possible on the nonperturbative
inputs, in particular, the light-cone wavefunctions of $\rho$ meson.
If confining LCSR computation to the present accuracy, the numerical
results signify that the heavy quark spin symmetry can kept well for
both the charge and the magnetic interactions of the negative-parity
heavy mesons with $\rho$ meson, but the heavy flavor symmetry
suffers from a violation of different degree in the two interaction
situations. To quantify size of the effects from the heavy flavor
symmetry breaking, we consider a ratio of the corresponding sum rule
results in the bottom and charm cases. We observe that whereas the
estimated ratios, for the charge interactions, are of a deviation of
about $20\%$ from 1, the resulting breaking effect is at a level of
a few percent in the cases of the magnetic interaction.

\section{Heavy Quark Limit and Determination of $\beta$ and $\lambda$ }

~~~~~In this section, we want to take a closer look at the behavior
of the strong couplings in the heavy quark limit, checking up the
consistency of the LCSR results with the predictions of heavy quark
spin symmetry, and providing an assessment of the low energy
effective parameters $\beta$ and $\lambda$.

The desired asymptotic forms are achievable from the corresponding
sum rules for the finite quark quark mass, by working explicitly out
$m_Q$ scaling behavior of the relevant parameters depending on the
heavy degree of freedom. To be specific, we need to substitute in
the sum rule results (18-21) the standard expansions of the $B
(B^*)$ meson mass $m_B (m_{B^*})$, decay constant $f_B(f_{B^*})$,
Borel parameter $M^2$ and threshold $s_0$. The former two are of the
following expansions in inverse $m_b$:
\begin{equation}
m_B (m_{B^*})=m_b+\Lambda+{\cal O}(1/m_b),~~~
f_B(f_{B^*})=F/\sqrt{m_b}+{\cal O}(1/m_b),
\end{equation}
with $\Lambda$ being the binding energy of the light degree of
freedom in the static $b$ quark chromomagnetic field, and $F$ a low
energy parameter. For the intrinsic parameters in the sum rules
$M^2$ and $s_0$, we need to rescale them as,
\begin{equation}
 M^2=2m_bT,~~~~~ s_0=m_b^2+2m_b\omega_0,
\end{equation}
with $T$ and $\omega_0$ being the $m_Q$ independent Borel variable
and threshold, respectively.

It turns out, with these expansions, that in the limit
$m_Q\rightarrow\infty$, the sum rules in (18-21), as desired, comply
precisely with the asymptotic relations $g_{BB\rho}=g_{B^*B^*\rho}$
and $f_{B^{\ast}B\rho}=f_{B^{\ast }B^*\rho }$, and so boil down to
the two dependent expressions. In consequence, the $m_Q$ scaling
behavior of the strong couplings are reproduced rightly and a
consistent result is obtained with the effective chiral lagrangian
approach, in the LCSR approach. Denoting the asymptotic forms of
$g_{BB\rho}(g_{B^*B^*\rho})$ and $f_{B^{\ast }B\rho}(f_{B^{\ast
}B^*\rho})$ by ${\cal G}_1$ and ${\cal G}_2$, respectively, the
resulting sum rules are of the following forms:
\begin{eqnarray}
F^2{\cal
G}_1&~=~&\frac{\sqrt{2}}{4}e^{\frac{\Lambda}{T}}\left[2\left(1-e^{-\frac{\omega_0}{T}}\right)f_\rho
m_\rho T\varphi_{\parallel}\left(1/2\right)\right.\non\\&&\left.+
m_\rho^2f_\rho^Th_{\parallel }^{\left(
s\right)}\left(1/2\right)-\frac{1}{8T}f_\rho
m_\rho^3\left(A\left(1/2\right)+8\widetilde{C}\left(
1/2\right)\right)\right],\label{28}
\end{eqnarray}
\begin{eqnarray}
F^2{\cal
G}_2&~=~&\frac{\sqrt{2}}{8}e^{\frac{\Lambda}{T}}\left[2\left(1-e^{-\frac{\omega_0}{T}}\right)f_\rho^TT\varphi
_{\perp}\left(1/2\right)+\frac{1}{2}f_\rho m_\rho
g_{\perp}^{(a)}\left(1/2\right)\right.\non\\&&-\left.\frac{1}{8T}f_\rho^Tm_\rho^2A_T\left(1/2\right)\right]
\label{29}.\end{eqnarray}

The numerical analysis of the above asymptotic sum rules can be made
using all the same procedure as in the finite heavy quark mass case.
In the first place, the binding energy $\Lambda$, as an important
input, requires to be fixed in a consistent way to reduce the
numerical uncertainty as much as possible. It is easily calculated
by taking the logarithmic derivative for one of (24) and (25) with
respect to the inverse Borel parameter $T$. The result from (24),
for instance, is $\Lambda=0.43\pm 0.15\gev$ with the Borel interval
$0.5\ll T\ll1.3\gev$ and threshold $\omega_0=1.3\pm 0.1\gev$. With
these inputs, we get $F^2{\cal G}_1=0.210\pm 0.031\gev^3$ and
$F^2{\cal G}_2=0.098\pm 0.013\gev^2$. The variations are depicted in
Fig.1 of the LCSR results with the Borel parameter $T$. In order to
have an assessment of the asymptotic couplings ${\cal G}_1$ and
${\cal G}_2$, we may use the determination without QCD radiative
corrections included \cite{Neubert}: $F=0.30\pm0.05 \gev^{3/2}$.
Instead of doing that, we prefer to directly substitute in (24) and
(25) the sum rule form for $F$ to make the numerical results free of
a large uncertainty, yielding ${\cal G}_1=2.36\pm 0.32$ and ${\cal
G}_2=1.09\pm 0.15 \gev^{-1}$, a result compatible with the sum rules
for the finite heavy quark mass.

Now we are in a position to determine the effective parameters
$\beta$ and $\lambda$. In the context of the effective chiral
lagrangian \cite{CHIRAL}, the related hadronic matrix elements obey,
at the leading order in the $1/m_{B^{(*)}}$, the following
parameterizations: \be \langle \bar
B^{0}(p)\rho^-(q,\epsilon)|B^{-}(p+q)\rangle=i\sqrt{2}M_B~\beta~
g_V~ \epsilon^*\cdot v \label{2} \en \be \langle \bar
B^{*0}(p,\eta)\rho^-(q,\epsilon)|B^{-}(p+q)\rangle=-i2\sqrt{2M_B~M_{B^*}}~\lambda~
g_V~\epsilon_{\mu\alpha\beta\gamma}\eta^{\ast\mu}q^{\alpha}
\epsilon^{\ast \beta}v^{\gamma} \label{3} \en \be \langle \bar
B^{*0}(p,\eta)\rho^-(q,\epsilon)&|&B^{*-}(p+q,\xi)\rangle =
-i\sqrt{2}M_{B^*}~\beta~g_V~\left(\eta^*\cdot\xi\right)
\left(\epsilon^*\cdot
v\right)\nonumber\\&&-i2\sqrt{2}M_{B^*}~\lambda~
g_V\left[\left(\eta^*\cdot\epsilon^*\right)\left(\xi\cdot
q\right)-\left(\xi\cdot\epsilon^*\right)\left(\eta^*\cdot
q\right)\right]\label{4}
  \en
where $g_V=m_{\rho}/f_{\pi}\approx 5.8$ \cite{CHIRAL} and $v$
indicates the velocity of the underlying heavy mesons. Confronting
the hadronic matrix elements in (2)-(4) with those in (26)-(28)
respectively, we have the following asymptotic relations:
\begin{eqnarray}
{\cal G}_1=\frac{\beta g_V}{2},~~~~~~~~{\cal G}_2=\frac{\lambda
g_V}{2}.
\end{eqnarray}
Using the above equation and sum rule results for ${\cal G}_1$ and
${\cal G}_2$, we get $\beta=0.81\pm0.11$ and $\lambda=0.38\pm
0.05\gev^{-1}$.

The authors of \cite{FSI2} give an estimate of the effective
couplings $\beta$ and $\lambda$. They consider the electromagnetic
transition of a heavy pseudoscalar meson and assume that the
hadronic matrix element of the light quark current is dominated by
the $\rho$, $\omega$,$\phi$ vector mesons. Then the current
conservation leads automatically to the result
$\beta=\frac{\sqrt{2}m_V}{g_Vf_V}\approx 0.9$. To order to make an
evaluation of the parameter $\lambda$, they adopt a combined use of
several different approaches. The prescription is to compute one of
the $B \to K^*$ form factors at the squared momentum transfer
$q^2=q_{max}^2$, using the effective chiral lagrangian and $B_s^*$
dominance model, respectively, and then to equate them for
extracting $\lambda$ which enters the result of the effective
theory. The pole model representation for the form factor is
determined by identifying, at $q^2=17\gev^2$, its result with the
corresponding theoretical prediction from the LCSR's and lattice
QCD. In such ways one gets $\lambda=0.57\gev^{-1}$. Also, it is
passible to extract $\lambda$ from the data on the $D\to K^*$ form
factor at the largest recoil, by extrapolating the form factor
derived at zero recoil in the effective chiral lagrangian approach
by means of the vector dominance \cite{CHIRAL}. The extracted
$\lambda$ is of a bit smaller central value:
$\lambda=0.41\gev^{-1}$. It is generally agreed that these existing
determinations of $\beta$ and $\lambda$ would be subject to a large
uncertainty, especially in the $\lambda$ case where a combined use
of several different approaches to the form factor would more or
less cause the inconsistency in calculation. Concentrating on the
central values, we find that QCD LCSR's predict, for the parameter
$\beta$, a numerical result nearly the same as the one of the pole
model. In the $\lambda$ case, a good numerical agreement is also
observed with the result extracted experimentally, whereas there is
a numerical deviation of about $-30\%$ from the one of \cite{FSI2}.
On the whole, our LCSR results for $\beta$ and $\lambda$ are
compatible with those of other approaches within errors.

\section{Summary}

~~~~~The strong interactions of the negative-parity heavy mesons
with $\rho$ meson may be described uniformly in the context of an
effective lagrangian observing $SU(2)$ invariance in the isospin
space. The established effective lagrangian contains four
independent coupling parameters, which characterize the dynamics of
strong interactions among the underlying meson fields. Using the QCD
LCSR method and recently updated model parameters for the light-cone
distribution amplitudes of $\rho$ meson, we have presented a
complete discussion on these couplings. Apart from an updated LCSR
result for $g_{BB\rho}$ and $f_{B^*B\rho}$, we give, among others, a
detailed LCSR estimate of $g_{B^*B^*\rho}$ and $f_{B^*B^*\rho}$,
about which little was known before. Situations of the charm mesons
are also inquired into in the same framework, which is especially
important for us to understand the FSI effects in B decays. A
systematic numerical discussion is made, including a detailed
physical interpretation on the sum rule results and a numerical
comparison with the LCSR computations using as inputs a model
wavefunction given earlier. Also, we examine asymptotic forms of the
LCSR results in the heavy quark limit. As shown explicitly, the LCSR
approach could reproduce rightly the $m_Q$ scaling behavior of the
physical quantities in question, and thus provide a consistent
calculation with the results of the heavy quark symmetry. This
would, needless to say, enhance considerably our confidence in
applying the LCSR method to do calculation of nonperturbative
quantities. Finally, we assess the low energy parameters $\beta$ and
$\lambda$ appearing in the corresponding effective chiral
lagrangian, and draw a numerical parallel between the present and
previous calculations.

The effective lagrangian approaches, using the present findings as
inputs and in conjunction with other nonperturbative methods, could
help to get a more knowledge of the long distance dynamics in heavy
meson weak decays. No doubt, this is beneficial to promote our
understanding of the standard model of particle physics.

\vspace{0.6 cm}{\Large\bf Acknowledgements} \vspace{0.3 cm}

This work was supported in part by the Natural Science Foundation of
China (NSFC) under No.10675089.

\newpage
\begin{figure}
\centerline{ \epsfxsize=12cm \epsfbox{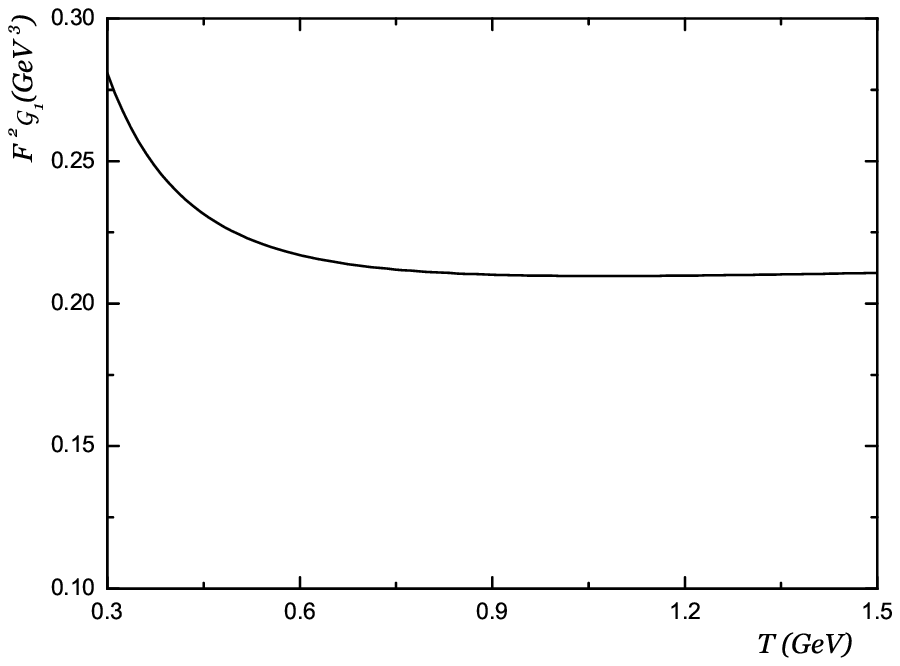}
}\vspace{-0.2cm} \centerline{(a)}\vspace{1cm} \centerline{
\epsfxsize=12cm \epsfbox{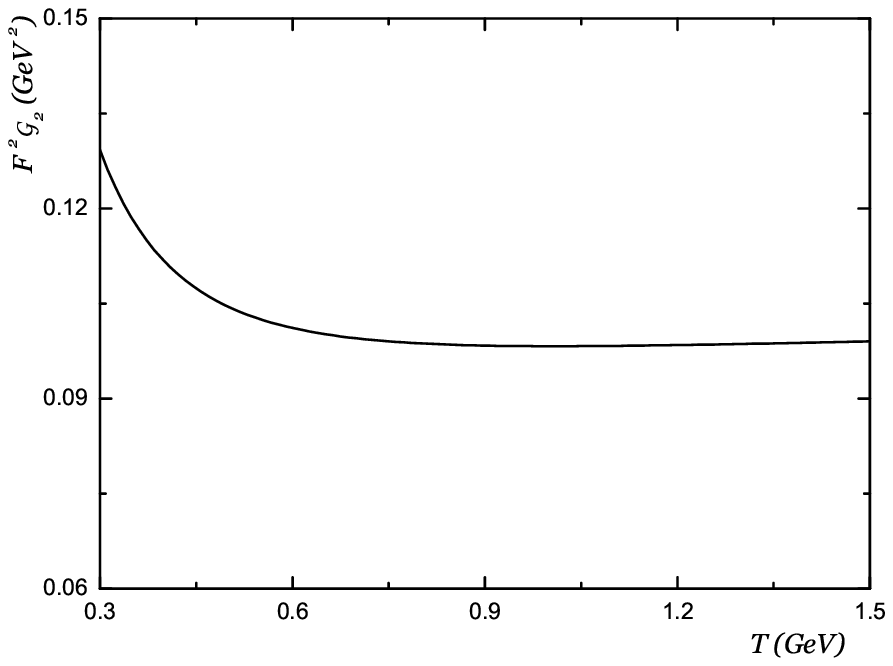} }\vspace{-0.2 cm}
\centerline{(b)}\vspace{0.8 cm} {Fig.1. The stability of the LCSR
results for $F^2{\cal G}_1$ (a), and $F^2{\cal G}_2$ (b), with
$\Lambda=0.43\gev$ and $\omega_0=1.3\gev$.}
\end{figure}

\end{document}